# Abnormal temperature dependence of mobility in conjugated polymer / nanocrystal composite: experiment and theory

YatingZhang[1,2]*, JianquanYao[1], Hoi Sing Kwok[2]
1 College of Precision Instrument and Opto-electronics Engineering, Tianjin University, CN-300072 Tianjin, P. R. China
2 Department of Electronic & Computer Engineering, Hong Kong University of Science and Technology, Clear Water Bay, Hong Kong SAR, P. R. China

Abstract:

Instead of normal non-Arrhenius relationship, the carrier mobility $\ln(\mu)$ v.s. $1/T^2$ showed abnormal dependence in an MEH-PPV / InP nanocrystal composite system that a critical temperature ($T_c$) behavior is prominent in temperature range of 233 K to 333 K. Here, in the model of variable range hopping theory, an analytical model is developed within a Gaussian trap distribution, which is successfully implemented on that phenomenon. The results show that $T_c$ becomes the transition temperature as long as trap-filling factor (FF) ~1, which means a transition point from Boltzmann to Fermi distribution. Furthermore, the model predicts an universal relationship of $\ln(\mu)$ on $1/T^2$ determined by FF in any disordered system with traps.

Carrier mobility is a very important parameter in electronic and optoelectronic devices [1-5]. As an advanced class of electronic and optoelectronic media, conjugated polymer / semiconductor nanocrystal composites have received extensive attention, due to potential technological applications in various optoelectric devices, such as organic light-emitting diodes (OLEDs), organic transistors, organic solar cells, and so on [6-8]. Compared with pristine polymer, there are positive and negative impacts on the mobility from embedded semiconductor nanocrystals. First of all, the guest states of the embedded semiconductor nanocrystals can induce traps who can promptly capture carriers and keep them localized over a long time , which is attributed to negative effect on mobility [9-12]. On the other hand, the interface of polymer and nanocrystals is beneficial to the separation of photo-induced excitons, thus the concentration of carrier (n) is larger than that in pristine polymer [13, 14]. Fermi level ($E_F$) rises up with the increase of n, and so as the mobility (μ) [3, 4, 15]. This positive contributions have been proved in many conjugated polymer / semiconductor nanocrystal composite systems, and μ increase at least an order of magnitude [16, 17]. For example, Andrew Watt *et al.* study the mobility of MEH (poly [2 - methoxy-5 - 2 - ethyl-hexyloxy] - p -phenylenevinylene) / PbS nanocrystal composite, and observed that carrier mobility enhanced from $2.83 \times 10^{-3}$ $cm^2V^{-1}s^{-1}$ to $9.58 \times 10^{-2}$ $cm^2V^{-1}s^{-1}$ [16]. Choudhury *et al* observed that the mobility in PVK (poly N-vinylcarbazole) / CdS nanocrystal composite increased with the increase of the concentration of CdS nanocrystals. In this case, it is believed nanocrystals do not introduce additional traps [18]. However, how do the traps introduced by nanocrystals affect the mobility has never been reported and become a question with profound interest.

Here, we designed a special composite MEH-PPV / InP nanocrystal to attempt answering this question. Surface states of InP are capable of introducing carrier traps, due to the proper energy level and high density. Considering the bottleneck of transportation in disordered system, thermal excitation and thermal relaxation of carriers [19], we studied temperature dependence of mobility by traditional time of flight (TOF) technique. The abnormal $1/T^2$ dependence of mobility is observed that ln(μ) linearly decreases with $1/T^2$ before and after a critical temperature ($T_c$). Through analysis based on a model of variable range hopping theory, we found that an often ignored parameter, trap-filling factor (FF), plays a critical role and determines which type of T dependence of ln(μ) works. There are three cases induced by FF (< 1; ~ 1; > 1) in total, among which a distribution transition of carrier can appear, when and only when FF approximately equals to 1. At this point a critical temperature emerged, as observed in our experiment. And then a universal T dependence of ln(μ) can be derived by the method in any disordered organic system with Gaussian traps determined by FF.

To explore whether nanocrystals would introduce additional traps, we carefully studied the energy structure of composites formed by different materials of the components, and found out that the traps formed by intrinsic energy states of embedded nanocrystals are too deep ( ≥ 1 eV ) for carriers to escape by thermal excitation. As a result, the impacts for these too deep traps cannot reflect themselves on curve μ(T) or μ(E). Therefore, the key point is to design and prepare a composite with *proper* trap depth (~ 0.3 eV). Here, we presented and prepared a special composite, MEH-PPV / InP nanocrystal composite whose energy states structure is shown in figure 1. The speciality lies in high density of surface states of InP nanocrystals who serve as the traps instead of the intrinsic states. In MEH-PPV, localized states distributed around 5.3 eV where positive charge carriers located; while the surface states of InP is quiet below 0.3 eV where carriers can relax and thermal escape, so they become traps more effective than the intrinsic ones of InP at 4.4 eV (depth of trap ~ 1 eV) as illustrated in the right section of figure 1. The preparation method is detailed in Ref.[20], and surface states is formed by In dangling bond [21].

The sandwich structured samples are prepared by spin casting the composite solution (or polymer solution) on Indium tin oxide (ITO), and then subsequently dried overnight under vacuum. Al as another electrode in ~ 1 μm thickness was then thermally evaporated over a shadow mask with an active area of ~ 1 $cm^2$. Schematic setup of TOF is shown in the left section of figure 1. In experiment, we use third harmonic wave from Nd:YAG pulse laser (wavelength: 355 nm, pulse width: 10 ns, repeat rate: 5 Hz) as excitation laser. When incident pulse laser penetrated ITO, the excitons will be generated in MEH-PPV; under the driving force of external electric field, the excitons separating into free negative and positive charge carriers, which are then drifted to the corresponding electrodes. When they reach the electrode, photo induced current can be measured in external circuit. By taking the transient voltage of resistor, transient current characteristics can be measured. During measurement, sample is fixed on a holder in a vacuum chamber located on a semiconductor thermostatic platform (LTD3-10). Agilent 54622A Oscilloscope (100MHz) takes the voltage across the resistor and puts digital signals into computer.

Dispersive transport is observed on transient current curves both for composite and polymer sample, as shown in the insert of Fig 2. The cross point of two straight lines is extracted as transit time ($\tau$). Then the mobility can be calculated by substituting $\tau$ into equation $\mu = d^2 / (\tau U)$, where d is the thickness and U is external bias voltage. Turning detection temperature, the mobility at different T can be collected. The hole mobility $\mu_H$, vs. $1/T^2$, are plotted in figure 2, where (a) for the reference of MEH-PPV and (b) for the nanocrystal composite.

Obviously, the linear dependence of $\ln(\mu)$ on $1/T^2$ is observed for MEH-PPV, termed as non-Arrhenius dependence [2]. It indicates that carrier transport is determined by thermal excitation and associated with the energetic disorder of organic system. This trend has been confirmed in numerous similar samples [2, 22-24], and hence is regarded as "normal behavior" here, whereas for the composite, abnormal behavior is evident. Two independent non-Arrhenius $\ln(\mu) \propto 1/T^2$ is broken by a critical temperature ($T_c = 283$ K), note that this abnormal mobility behavior in our composite structure is highly repeatable.

In order to understand $T_c$, we developed an analytic model of charge carrier transport in a disordered system, as following.

In MEH-PPV, holes transport via hopping within the energy level of 5.3 eV, which is made up of many localized states. Setting zero energy locates at this energy level, the distributed states are described by Gaussian function as[24]

$$g(E) = \frac{N}{\sqrt{2\pi}\sigma} \exp\left[-\left(\frac{E}{\sigma}\right)^2\right] \quad (1)$$

Where N ($\sim 10^{22}$ cm$^{-3}$) is total density of localized states, $\sigma$ is the width of Gaussian distribution usually in orders of $\sim 0.1$ eV [4, 5, 25]. Note that the reference zero energy is set to energy level 5.3 eV, and also the center of state distribution. The density of states (DOS) is normalized to $\int_{-\infty}^{\infty} g(E)dE = N$. Similarly, one can write out the DOS of the composite, with trap states distributed around center energy $E_{Tr}$, and the width of $\sigma_{Tr}$ [9],

$$g(E) = \frac{N}{\sqrt{2\pi}\sigma}\exp\left[-\left(\frac{E}{\sigma}\right)^2\right] + \frac{N_{Tr}}{\sqrt{2\pi}\sigma_{Tr}}\exp\left[-\left(\frac{E-E_{Tr}}{\sigma_{Tr}}\right)^2\right] \quad (2)$$

where $E_{Tr}$ represents the center energy of surface states of InP naocrystals. In our experiment, it equals to 0.3 eV. Total DOS has to be normalized to $\int_{-\infty}^{\infty} g(E)dE = N + N_{Tr}$, where $N_{Tr}$ is the total density of surface states of InP nanocrystals, and lies in the range of 0.1N ~ 0.001N depending on the mixing ratio of polymer and InP nanocrystals. According to Mill-Abrahams rate model, charge carriers jump from starting state of energy $E_s$ to target state of energy $E_t$ over distance $r_{st}$ [9]

$$\gamma_{st} = \gamma_0 \exp\left(-2\frac{r_{st}}{a}\right)\exp\left[-\frac{E_t - E_s + |E_t - E_s|}{2kT}\right] \quad (3)$$

where $\gamma_0$ is attempt-to-frequency, in the order of $10^{12}$ Hz, $a$ is the localization radius or localization length [3, 9, 26]. In thermal equilibrium, the DOS occupied by carriers can be described by Fermi function $f(E, E_F) = \frac{1}{1+\exp(E-E_F)}$ with the Fermi energy level $E_F$ determined by condition

$$\int_{-\infty}^{\infty} f(E, E_F)g(E)dE = n \quad (4)$$

where n is the concentration of carriers. Eqs. (1), (3) and (4) formulate a self-consistent theoretical model with three dimensionless parameters: $\sigma/kT$, $Na^3$ and $n/N$. This is the common situation without traps; on the other hand, Eqs. (2), (3) and (4) formulate a theoretical model with traps. Therefore, we must introduce an additional important condition to describe the degree of carriers filling in the DOS of trap states: Trap filling factor (FF = $n/N_{Tr}$). Using the concept of FF, three cases can be clearly separated.

The first condition is: FF >> 1, which means that the states of traps are fully filled by carriers, and those of polymer are partially filled. As well known, charge carriers dive in energy unlimitedly in course of time in an empty system at kT < σ [3], until carriers arrive at the vicinity of the Fermi level $E_F$. Then, thermal excitation induces carriers to perform hops to the states in the vicinity of a very import energy $E_T$. In the states around $E_T$ carriers fall into the states that have lower energies. This process near the $E_T$ resembles a multiple-trapping process, where $E_T$ is called transport energy (TE), representing the mobility edge [1, 15, 26]. Since the upward hopping from Fermi level to TE dominates the charge transport, this case is very close to that in absence of the trap states.

Here, we discuss FF = 3 as an example illustrated in upper-left section of figure 3. Black line denotes the distribution of states, and light green area represents states occupied by carriers. The boundary of occupied and unoccupied states by carriers becomes $E_F$. Dark green line denotes TE. The arrow represents the dominated upward hopping. With rise of temperature, $E_F$ decreases and $E_T$ increases by solving Eq. (4) and (8), respectively. When calculating, we used parameters as followes, $E_{Tr}$ = -0.3 eV, $\sigma_{Tr}$ = 0.03 eV, $\sigma$ = 0.08 eV, N = $1\times10^{22}$ cm$^{-3}$, $\sigma_{Tr}$ = 0.03 eV, $N_{Tr}$ = 0.01N, and FF = 3, according to Ref. [3], [9] and our experiment.

Using method in Refs.[3, 15], the mobility μ(T) can be expressed as:

$$\mu = \frac{e}{kT} R^2(E_T) \langle t \rangle^{-1} \tag{5}$$

$$R(E_T) = \left\{ \frac{4\pi}{3} B^{-1} \int_{-\infty}^{E_T} E_T g(E)[1 - f(E, E_F)] dE \right\}^{-1/3} \tag{6}$$

$$\langle t \rangle = \gamma_0^{-1} \frac{\int_{-\infty}^{E_T} \exp\left[ \frac{2R(E_T)}{a} + \frac{E - E_T}{kT} \right] g(E)[1 - f(E, E_F)] dE}{\int_{-\infty}^{E_T} g(E)[1 - f(E, E_F)] dE} \tag{7}$$

Where $E_T$ is determined by equation

$$\frac{2}{3}\left(\frac{4\pi}{3B}\right)^{-1/3} \frac{kT}{a} \left[ \int_{-\infty}^{E_T} g(E)[1 - f(E, E_F)] dE \right]^{-4/3} [1 - f(E, E_F)] g(E_T) = 1 \tag{8}$$

When calculating the carrier mobility, we also take into account the percolation nature of hopping conduction, namely, that in order to provide an infinite percolation cluster of connected sites, in average B ~ 2.7 [27]. Substituting Eqs. (6) and (7) into Eq. (5), one can obtain the carrier mobility

$$\mu = \gamma_0 \frac{e}{kT} \frac{3B}{4\pi R(E_T) n_t} \exp\left[ -\frac{2}{a} R(E_T) - \frac{E_T - E_F}{kT} \right] \tag{9}$$

Where $n_t$ is determined by

$$n_t = \int_{-\infty}^{E_T} g(E) f(E, E_F) dE \tag{10}$$

Since TE is very close to the reference energy [3, 15], carrier concentration below TE $n_t$ approximately equals to total concentration n. So, $n_t$ can be replaced by n in Eq. (9) when calculation. Using Eqs. (2), (4), (6), (8) and (9), we calculate the mobility and plot $\ln(\mu)$ vs. $1/T^2$ in figure 4, with additional parameters $1/a = 6.25$ nm$^{-1}$, B ~ 2.7, according to Ref [3], [9] and the experiment.

The second case is when FF << 1. In low-concentration regime, Fermi distribution reduces to Boltzmann approximation as $f(E, E_F) = 1/\{1 + \exp[(E - E_F)/kT]\} \approx \exp(E_F/kT)\exp(-E/kT)$. Substituting it into Eqs. (9) and (10), one can cancel n, which means the mobility loses it concentration dependent property. Then, carriers thermal relax to equilibrium energy ($E_{eq}$), replacing Fermi level and becoming the new start point of upward hopping, which is expressed as

$$E_{eq} = \int_{-\infty}^{\infty} E g(E) \exp(-\frac{E}{kT}) dE \Big/ \int_{-\infty}^{\infty} g(E) \exp(-\frac{E}{kT}) dE$$

$$= -\frac{N\frac{\sigma^2}{kT}\exp\left[\frac{1}{2}\left(\frac{\sigma}{kT}\right)^2\right] + N_{Tr}\left(-E_{Tr} + \frac{\sigma_{Tr}^2}{kT}\right)\exp\left[\frac{1}{2}\left(\frac{\sigma_{Tr}}{kT}\right)^2 - \frac{E_{Tr}}{kT}\right]}{N\exp\left[\frac{1}{2}\left(\frac{\sigma}{kT}\right)^2\right] + N_{Tr}\exp\left[\frac{1}{2}\left(\frac{\sigma_{Tr}}{kT}\right)^2 - \frac{E_{Tr}}{kT}\right]} \quad (11)$$

Due to $N_{Tr} \ll N$ and $\exp(-E_{Tr}/kT) \ll 1$, the first term is considerably greater than the second term both in numerator and in denominator, therefore, $E_{eq} \approx N\frac{\sigma^2}{kT}\exp\left[\frac{1}{2}\left(\frac{\sigma}{kT}\right)^2\right] \Big/ N\exp\left[\frac{1}{2}\left(\frac{\sigma}{kT}\right)^2\right] = \frac{\sigma^2}{kT}$.

The lower-left section of figure 3 shows the situation of FF = 0.01 as an example. In the same distribution of states, when concentration of carrier reduces, carriers relax to $E_{eq}$ in width of σ, red area denotes states occupied by carriers. Then arrow marks the new upward hopping. These two important energy levels ($E_T$ and $E_{eq}$) both increase with temperature, obviously. Hence, T dependence of $\ln(\mu)$ are calculated and plotted in figure 4 with the same values of parameters except for FF.

The TE is very close to reference "0" energy, and changes little with temperatue than other two levels.; whereas, $E_F$ and $E_F$ show opposite tendency. If value of FF is too large or small, as cases 1 or 2, they cannot intersect in the studied temperature range. It means carriers follow and hold on either Fermi distribution (at high concentration) or Boltzmann distribution (at low concentration), and never change with temperature.

The green and black lines in figure 4 denote the first and second cases, respectively. In the same disordered system, temperature dependence of mobility show different characters depending on FF. For Fermi distribution (FF >> 1), the Fermi level $E_F$ is the starting point of thermal excited hopping; whereas, for Boltzmann distribution (FF << 1), the equilibrium energy $E_{eq}$ alternatively becomes the starting point of hopping. Since $E_{eq}$ is more sensitive to T than $E_F$, mobility under Boltzmann distribution is more sensitive to T than that under Fermi distribution. That is the reason black curve is steeper than the green one in figure 4.

The third case is the most complex one and will show an abnormal property when FF ~ 1. In this regime, the concentration $n = \int_{-\infty}^{\infty} f(E, E_F) g(E) dE$ is comparable to the density of trap states $N_{Tr} = \int_{-\infty}^{E_{Tr}+2\sigma_{Tr}} g(E) dE$, $\int_{-\infty}^{\infty} f(E, E_F) g(E) dE \approx \int_{-\infty}^{E_{Tr}+2\sigma_{Tr}} g(E) dE$, so that Fermi level $E_F$ is pinning at the overlapping region of two Gaussian distributed states. As a result, $E_F$ and $E_{eq}$ intersect at some temperature, as right section of figure 3. When $E_F$ lies below $E_{eq}$, i.e. $E_F \leq E_{eq}$, Boltzmann distribution holds; whereas when $E_F$ lies above $E_{eq}$, i.e. $E_F > E_{eq}(T)$, Boltzmann distribution is invalid, and Fermi distribution holds. And then, with rise of T, the start point of upward hopping shifts from $E_{eq}$ to $E_F$ at the crossing point of T. As a consequence, the mobility $\ln(\mu)$ versus $1/T^2$ shows a turning point, or a critical temperature, which separates the two kinds of transportation. Using Eqs. (6), (9) and (10), one can calculate the $\ln(\mu)$ versus $1/T^2$, which is plotted in figure 4 with the same values of parameters in above two case except for FF. Therefore, crosspoint of $E_F$ and $E_{eq}$ becomes the critical temperature ($T_c$), 282.4 K is calculated by the model, well matched with 283 K in experiment. Moreover, it is perfectly explained why other researchers did not observed that phenomenon, i.e. in the Choudhury *et al*'s paper [18], $T_c$ phenomenon is not shown in T dependence of $\ln(\mu)$ since FF >> 1, and the transition of carriers distribution ocurrs only when the density of effective trap states approximately equals to the concentration of carriers, or FF ~ 1. With such method, one can calculate the mobility versus temperature conveniently.

In summary, by theoretical model analysis we found that there are three distinguish behaviors of T dependence of mobility, determined by FF. When FF >> 1, representing traps are fully filled, carriers follow Fermi distribution, and it is similar to the case without traps. When FF << 1, representing very low carrier concentration, Boltzmann distribution of carriers is applied. The carriers will thermally relax to equilibrium energy ($E_{eq}$), which becomes the starting point of the thermal excitation replacing the Fermi level in the first case; meanwhile the mobility lost concentration dependent property. When FF ~ 1, representing the concentration of carriers is approximately equals to the density of trap states, a turning point of $\ln(\mu)$ versus $1/T^2$ appears, as observed in the experiment, due to transition from Fermi distribution to Boltzmann distribution of carriers. This work reveals the influence of filling effects on temperature dependence of mobility in organic disordered system, and an universal $\ln(\mu)$ versus $1/T^2$ is developed.

We thank Prof. Sheng Chu for revising our manuscript in English. Financial support from National Natural Science Foundation of China (NSFC), project No.61271066, and Independent Innovation Foundation of Tianjin University (IIFTJU) No.020-60302070 is gratefully acknowledged.

References:

[1] L. Li, G. Meller, and H. Kosina, Applied Physics Letters 92 (2008)

[2] N. I. Craciun, J. Wildeman, and P. W. M. Blom, Physical Review Letters 100 (2008) 056601.

[3] J. O. Oelerich, D. Huemmer, and S. D. Baranovskii, Physical Review Letters 108 (2012)

[4] W. F. Pasveer, J. Cottaar, C. Tanase, R. Coehoorn, P. A. Bobbert, P. W. M. Blom, D. M. de Leeuw, and M. A. J. Michels, Physical Review Letters 94 (2005) 206601.

[5] C. Tanase, E. J. Meijer, P. W. M. Blom, and D. M. de Leeuw, Physical Review Letters 91 (2003) 216601.

[6] V. L. Colvin, M. C. Schlamp, and A. P. Alivisatos, Nature 370 (1994) 354.

[7] Y. Xu, M. Benwadih, R. Gwoziecki, R. Coppard, T. Minari, C. Liu, K. Tsukagoshi, J. Chroboczek, F. Balestra, and G. Ghibaudo, Journal of Applied Physics 110 (2011)

[8] N. Fuke, L. B. Hoch, A. Y. Koposov, V. W. Manner, D. J. Werder, A. Fukui, N. Koide, H. Katayama, and M. Sykora, Acs Nano 4 (2010) 6377.

[9] V. I. Arkhipov, P. Heremans, E. V. Emelianova, G. J. Adriaenssens, and H. Bässler, Journal of Physics: Condensed Matter 14 (2002) 9899.

[10] Y. Y. Yimer, P. A. Bobbert, and R. Coehoorn, Journal of Physics: Condensed Matter 20 (2008) 335204.

[11] I. I. Fishchuk, A. K. Kadashchuk, A. Vakhnin, Y. Korosko, H. Bässler, B. Souharce, and U. Scherf, Physical Review B 73 (2006) 115210.

[12] V. I. Arkhipov, E. V. Emelianova, P. Heremans, and H. Bässler, Physical Review B 72 (2005) 235202.

[13] H. van Eersel, R. A. J. Janssen, and M. Kemerink, Advanced Functional Materials 22 (2012) 2700.

[14] D. Pitsa and M. G. Danikas, Nano 6 (2011) 497.

[15] J. O. Oelerich, D. Huemmer, M. Weseloh, and S. D. Baranovskii, Applied Physics Letters 97 (2010)

[16] A. Watt, T. Eichmann, H. Rubinsztein-Dunlop, and P. Meredith, Applied Physics Letters 87 (2005)

[17] K. R. Choudhury, M. Samoc, A. Patra, and P. N. Prasad, Journal of Physical Chemistry B 108 (2004) 1556.

[18] K. R. Choudhury, J. G. Winiarz, M. Samoc, and P. N. Prasad, Applied Physics Letters 82 (2003) 406.

[19] F. Laquai, G. Wegner, and H. Baessler, Philosophical Transactions of the Royal Society a-Mathematical Physical and Engineering Sciences 365 (2007) 1473.

[20] Y. J. Zhang Yating, Advanced Materials Research 531 (2012) 31.

[21] O. I. Mićić, J. Sprague, Z. Lu, and A. J. Nozik, Applied Physics Letters 68 (1996) 3150.

[22] E. Lebedev, T. Dittrich, V. Petrova-Koch, S. Karg3, and W. Brütting, Applied Physics Letters 71 (1997) 2686.

[23] S. V. Rakhmanova and E. M. Conwell, Applied Physics Letters 76 (2000) 3822.

[24] H. Bässler, physica status solidi (b) 175 (1993) 15.

[25]     P. M. Borsenberger, R. Richert, and H. Bässler, Physical Review B 47 (1993) 4289.

[26]     V. I. Arkhipov, E. V. Emelianova, and G. J. Adriaenssens, Physical Review B 64 (2001) 125125.

[27]     O. Rubel, S. D. Baranovskii, P. Thomas, and S. Yamasaki, Physical Review B 69 (2004) 014206.

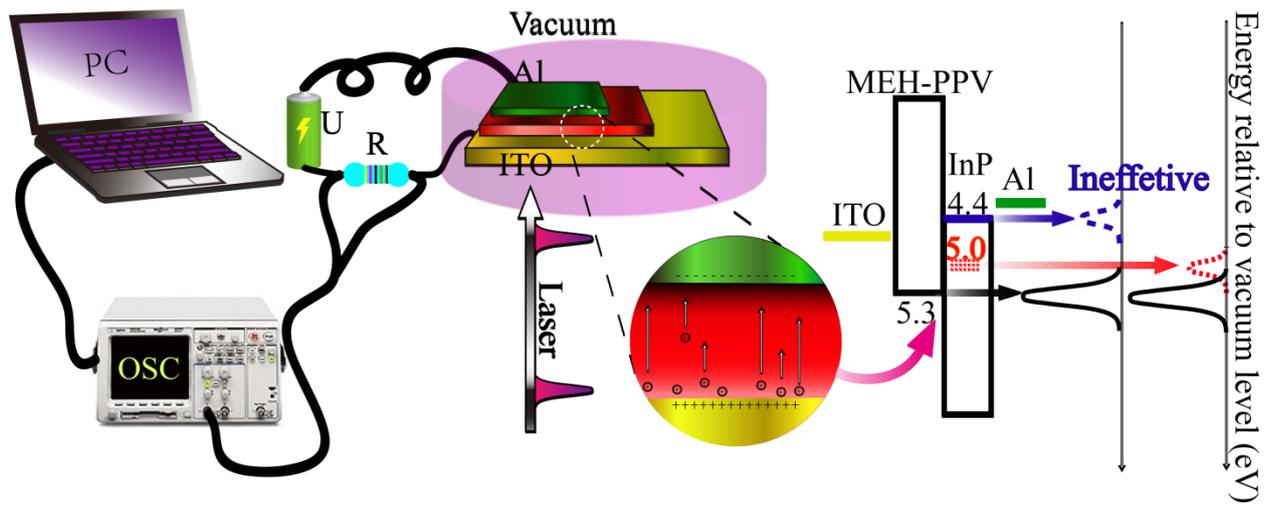

FIG.1.(a) Schematic structure of the energy level relative to vacuum level, (b)Schematic diagram of TOF setup.

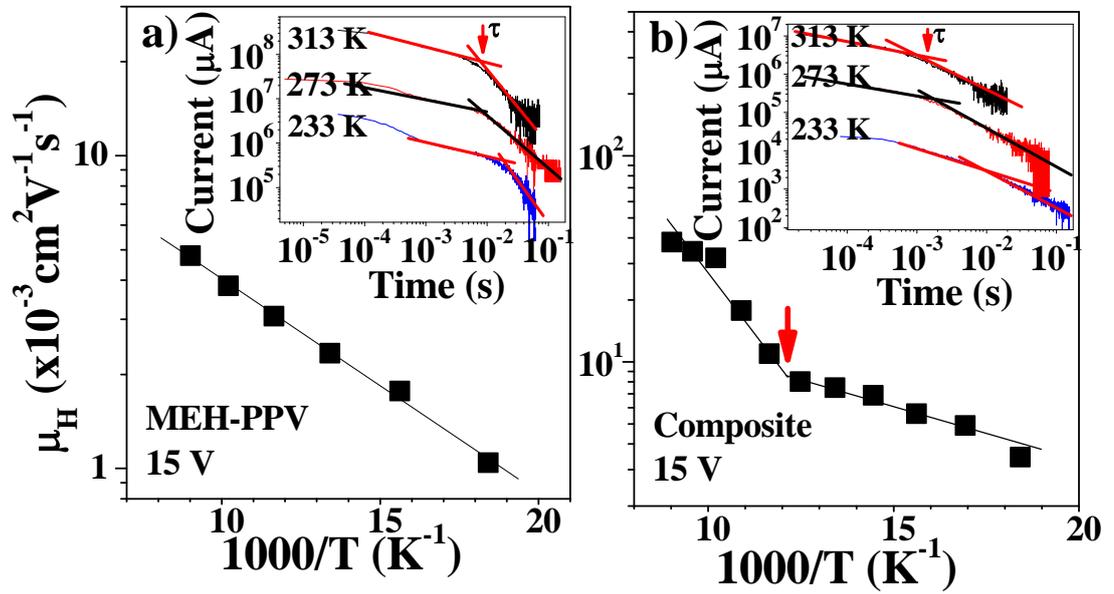

FIG. 2. Temperature dependent hole mobility in MEH-PPV (a) and composite (b), under biased voltage 15V, the inset is corresponding transient current under different temperature.

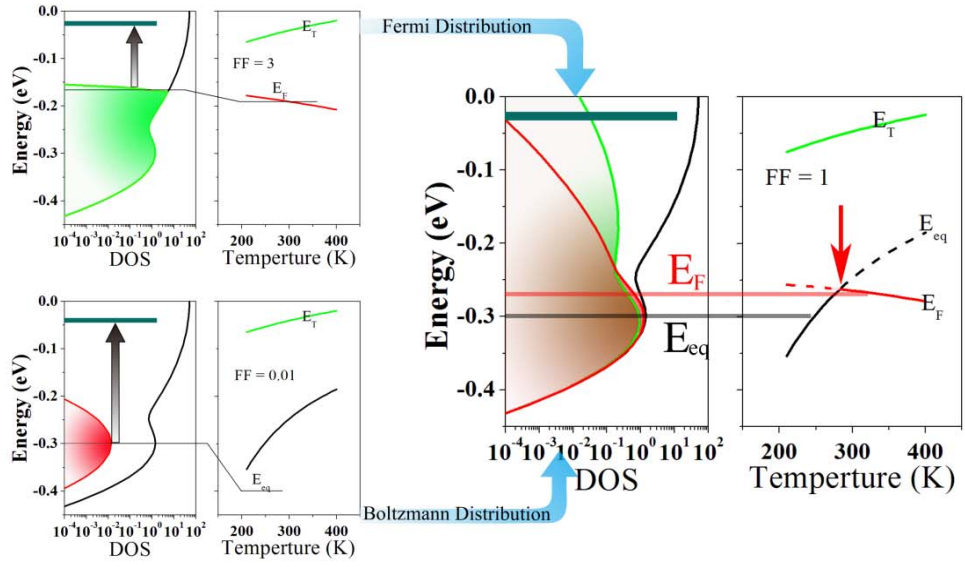

FIG. 3.(a) DOS of the composite, (b) three important energy levels $E_T$, $E_F$ and $E_{eq}$, vary with temperature at FF = 0.01, 1 and 3, respectively.

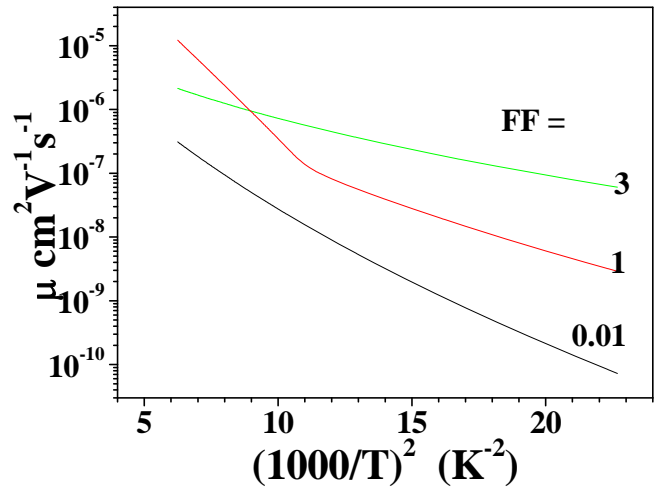

FIG.4. Calculated mobility versus temperature at FF = 3, 1 and 0.01, respectively.